\documentclass[prb,twocolumn,showpacs,preprintnumbers,amsmath,amssymb]{revtex4-1}
\usepackage{graphicx}
\usepackage{epsfig}
\def\olcite#1{[\onlinecite{#1}]}
\def\secref#1{section \ref{#1}}
\def\figref#1{figure \ref{#1}}
\def\Real{{\rm Re}\,}

\begin{document}
\title{Effect of normal current on kinematic vortices}
\author{P.~Lipavsk\'y$^{1,2}$, Pei-Jen Lin$^3$, Peter~Matlock$^4$ and A.~Elmurodov$^1$}
\affiliation{$^1$ Faculty of Mathematics and Physics, Charles University, 
Ke Karlovu 3, 12116 Prague 2, Czech Republic}
\affiliation{$^2$Institute of Physics, Academy of Sciences, 
Cukrovarnick\'a 10, 16253 Prague 6, Czech Republic}
\affiliation{$^3$Department of Physics, Old Dominion University, Norfolk, Virginia, USA}
\affiliation{$^4$Research Department, Universal Analytics Inc., Airdrie, AB, Canada}


\begin{abstract}
Within the framework of time-dependent Ginzburg-Landau theory, we
discuss an effect of the non-magnetic interaction between the normal
current and the supercurrent in the phase-slip regime. The correction
due to the current-current interaction is shown to have a transient
character so that it contributes only as a system evolves. Numerical
analyses for thin layers with no magnetic feedback show that the
largest contribution of the current-current interaction appears near
sample edges, where kinematic vortices reach maximum velocity.
\end{abstract}
\maketitle

\section{Introduction} 
Although time-dependent Ginzburg-Landau (TDGL) theory 
is justified only for slowly evolving systems, it provides 
qualitatively correct description of such phenomena
as the ultrafast propagation of magnetic flux dendrites
\cite{Biehler05,Vestgarden11}, fast kinematic vortices
\cite{Andronov93,Berdiyorov09}
or accelerated vortex motion involving vortex-antivortex
annihilation\cite{Sardella09}. Of course, in these fast
processes the quasiparticles cannot achieve a local 
equilibrium distribution, meaning that non-equilibrium corrections
to the TDGL theory become important. This was demonstrated 
by Vodolazov and Peeters\cite{Vodolazov10} who found a 
large deformation of the gap profile at the phase-slip 
centre in the case of a slow relaxation of quasiparticles. 

Vodolazov and Peeters have assumed isotropic distribution 
of quasiparticles (valid in the dirty limit) and carefully
treated the energy distribution using two coupled kinetic 
equations for longitudinal and transverse branches.
Their approach applies for a finite gap, the 
time derivative of which acts as a force driving 
quasiparticles out of equilibrium. 

Here we shall discuss a complementary correction (valid in the pure
limit) which takes into account a direction-dependent perturbation of
the momentum distribution of quasiparticles; such a perturbation appears
as the normal current is created. The correction to the TDGL
equation is found to be proportional to the scalar product of the
normal current and the supercurrent.\cite{LL08}

\subsection{Normal current in a superconductor}
Superconductors with freely-moving Abrikosov vortices or 
propagating dendrites have a finite resistivity; an 
electric field ${\bf E'}$ thus develops in them as the 
current is driven through. This electric field generates 
a normal current 
\begin{align}
{\bf J}_{N}=\sigma_{N}{\bf E'},
\label{e1}%
\end{align}
which adds to the supercurrent ${\bf J}_{S}$. In the TDGL theory the
normal current and the supercurrent interact only indirectly via the
magnetic field. The absence of any direct interaction between these
two currents in the TDGL theory is not disturbing, because it is in
agreement with an intuitive picture based on the two-fluid model of a
superconductor: taking the condensate as an independent fluid one
expects it not to interact with the underlying crystal including its
normal electrons.

The absence of interaction between normal current and supercurrent is
also supported by microscopic theories in the dirty
limit.\cite{SS75,KW78,Hu80,PWVHW82} These approaches, however, cannot
be used to discuss the current-current interaction. To obtain
practical equations, authors employ the isotropic
approximation\cite{KW78,PWVHW82,Vodolazov10} in some cases assuming in
addition local equilibrium\cite{SS75}. The isotropic distribution
corresponds to the zero normal current, therefore any effect of the
normal current on a formation of superconducting gap is lost in this
approximation.

The effect of the normal current on the gap has been derived 
in \olcite{LL08} from the Thouless criterion
\cite{Thouless60} adapted to non-equilibrium Green functions. 
In \olcite{LL09} it was applied to the FIR conductivity 
of the Abrikosov vortex lattice and shown to explain a decrease 
of the real part of conductivity below the critical temperature,
experimentally observed with ultrafast spectroscopy
\cite{ISIFK09}, while the TDGL theory predicts a small increase. 
Since the microscopic derivation is lengthy and technically 
demanding, in the appendix we provide a simple 
derivation of the interaction of the condensate with the normal 
current using purely phenomenological arguments.

\subsection{Plan of paper}
The paper is organized as follows. In \secref{GLFK} we introduce the
floating-kernel approximation, which is the TDGL theory extended by
the interaction between the normal current and the supercurrent. In
\secref{EF} we show that this correction is of transient nature, being
zero in any steady regime. To this end in \secref{EP} we perform a
gauge transformation to express conveniently the interaction of the
normal current and supercurrent in terms of time derivatives of the
vector and scalar potentials. Consequently, this correction can be
described in terms of effective magnetic and electric fields as shown
in \secref{EEMF}. In \secref{FV} we apply the theory to the phase-slip
regime in a strip made of thin superconducting layers with negligible
magnetic feedback.  After rescaling so all quantities are
dimensionless, in \secref{FV} we present results of a numerical
simulation to demonstrate how the current-current interaction
influences fast kinematic vortices in the phase-slip regime.  Section
\ref{conc} contains concluding discussion. In the appendix we indicate
why the TDGL theory violates the longitudinal $f$-sum rule and show
that addressing this problem with an intuitive two-fluid correction
leads to the floating-kernel approximation.

\section{Floating-kernel approximation}
\label{GLFK}
Here we write down a closed set of equations forming the theory we
term the floating-kernel (FK) approximation. This becomes identical to
TDGL theory in the coordinate system floating with normal electrons,
because the normal current vanishes in this reference frame.  When the
normal current is accelerated, the condensate experiences inertial
forces absent in the laboratory TDGL theory.

\subsection{Order parameter}
In the presence of normal current the time-evolution of the
order parameter is described by\cite{LL08,LL09}
\begin{align}
&{1\over 2m^*}\left(-i\hbar\nabla-{e^*\over c}{\bf A}-
{m^*\over en}{\bf J}_{N}\right)^2\psi +\alpha\psi+
\beta\left|\psi\right|^2\psi
\nonumber\\  
&~~~=-
{\Gamma\over\sqrt{1+C^2|\psi|^2}}\left({\partial\over\partial t}+{i\over\hbar}e^*\phi+{C^2\over 2}
{\partial|\psi|^2\over\partial t}\right)\psi.
\label{e2}%
\end{align}
The right hand side has been derived by Kramer and Watts-Tobin three
decades ago.\cite{KW78,WKK81} Terms $\alpha\psi$ and
$\beta\left|\psi\right|^2\psi$ are a standard part of
Ginzburg-Landau theory. The kinetic energy in the left hand side has
been obtained only recently.\cite{LL08} It applies to pure
superconductors, when the Cooperon mass equals twice the electron
mass, $m^*=2m$.  In the appendix this normal-current correction is
deduced from the longitudinal $f$-sum rule. The TDGL
equation\cite{KW78} obtains setting ${\bf J}_{N}=0$.

Away from the critical line the phase and amplitude relax with
different rates. The parameter controlling this difference is
$C=2\tau_{\rm in}\Delta_0/ (\hbar|\psi_0|^2)$, where $\Delta_0$ and
$\psi_0$ are values of the BCS gap and GL function at given
temperature in the absence of currents. Since in pure superconductors
$\tau_{\rm in} \Delta_{\rm BCS}\gg\hbar$, the correction $C^2|\psi|^2$ can
be large under realistic conditions. We use this relaxation rate in
the numerical example. Close to the critical line this correction
vanishes and one can use a simpler theory corresponding to the limit
$C\to 0$ of equation \eqref{e2}.

Our major concern will be the contribution of the normal current, the ${\bf J}_{N}$-term. 
In \eqref{e2} the kinetic energy depends on the 
difference between the velocity of the condensate 
\begin{align}
{\bf v}_{S}=
{1\over m^*}\left(\hbar\nabla\chi-{e^*\over c}{\bf A}\right),
\label{e3}%
\end{align}
where $\psi=|\psi|{\rm e}^{i\chi}$,
and the mean velocity of normal electrons 
\begin{align}
{\bf v}_{N}={1\over en}{\bf J}_{N}.
\label{e4}%
\end{align}
The first term of \eqref{e2} is thus the kinetic energy which must be
yielded by a pair of normal electrons in order to join the condensate,
in the reference frame floating with normal electrons.  To distinguish
the theory based on the velocity difference from the standard TDGL
theory, we call equation \eqref{e2} the floating-kernel
approximation.\cite{LL09}

\subsection{Two-fluid picture of current}
The derivative of the kinetic energy with respect to vector potential
${\bf A}$ defines the current operator. The correction to the normal
current thus also appears in the supercurrent 
\begin{align}
{\bf J}_{S}&={e^*\over m^*}{\rm Re}~\bar\psi
\left(-i\hbar\nabla-{e^*\over c}{\bf A}-
{m^*\over en}{\bf J}_{N}\right)\psi
\nonumber\\
&=e^*n_{S}\left({\bf v}_{S}-{\bf v}_{N}\right),
\label{e5}%
\end{align}
with $n_{S}=|\psi|^2$ being the density of Cooper pairs or the
condensate density. 
That this supercurrent depends on the relative velocity of the
condensate with respect to the normal background is
desirable. According to Ohm's law \eqref{e1} all electrons move with
the normal velocity ${\bf v}_{N}$. If a superconducting fraction moves
with a different velocity ${\bf v}_{S}$, we must add the
difference.

One may quit the picture of relative motion and rearrange the 
total current in the spirit of the two-fluid model
\begin{align}
{\bf J}&={\bf J}_{S}+{\bf J}_{N}
\nonumber\\
&=e^*n_{S}\left({\bf v}_{S}-{\bf v}_{N}\right)+en{\bf v}_{N}
\nonumber\\
&=e^*n_{S}{\bf v}_{S}+\left(en-e^*n_{S}\right){\bf v}_{N}
\nonumber\\
&={e^*\over m^*}{\rm Re}~\bar\psi
\left(-i\hbar\nabla-{e^*\over c}{\bf A}\right)\psi+{\bf J}_{N}
\left(1-{2|\psi|^2\over n}\right).
\label{e6}%
\end{align}
In this rearrangement the supercurrent has the condensate velocity
${\bf v}_{S}$. The correction term has become a part of the normal 
current, where it reduces the density of electrons to the fraction 
of normal electrons. We have used that Cooperon charge is twice the 
electron charge $e^*=2e$. 

The necessity to reduce the normal current to the normal fraction 
follows from the longitudinal $f$-sum rule.
In the appendix we show that in 
order to achieve a consistent theory formulated in terms of a free energy, the 
reduced normal current must be accompanied by changes in the free
energy which lead to the floating-kernel approximation. 

\subsection{Scalar and vector potential}
The vector and the scalar 
potentials ${\bf A}$ and $\phi$ yield the electric field 
\begin{align}
{\bf E'}=-{1\over c}
{\partial{\bf A}\over\partial t}-\nabla\phi. 
\label{e7}%
\end{align}
In some applications one should keep in mind that $\phi$ is thought of
as a local electrochemical potential, and not the electrostatic
potential. The vector ${\bf E'}$ is thus the driving force per
electron rather than the Maxwell electric field. Following the
notation of Josephson we write ${\bf E'}$ rather than ${\bf E}$ as a
reminder of this distinction. As is usual in the theory of
superconductivity we call ${\bf E'}$ the electric field for brevity.

Although the system has non-zero scalar potential, deviations 
from charge neutrality are so small that one can neglect 
them using the continuity equation in its stationary form 
$\nabla\cdot{\bf J}=0$. Substituting for the normal current
from \eqref{e1}, one finds the usual condition for the
potential
\begin{align}
\sigma_{N}\nabla^2\phi=\nabla\cdot{\bf J}_{S}.
\label{e8}%
\end{align}
We have used $\nabla\cdot{\bf A}=0$ and assumed a homogeneous sample;
$\nabla\sigma_{N}=0$. To evaluate the vector potential
we need the Maxwell equation
\begin{align}
\nabla^2{\bf A}=-\mu_0\left({\bf J}_{S}+{\bf J}_{N}\right)
\label{e9}%
\end{align}
which is also in the stationary approximation to be consistent with
the continuity equation. The set of equations
(\ref{e1}-\ref{e2}), \eqref{e5}, and (\ref{e7}-\ref{e9}) is 
closed.  

\section{Transient nature of the interaction of the normal current with the supercurrent}\label{EF}
An overlap of the normal current and supercurrent appears at the
conversion layer at the junction of the superconductor to a normal
lead. Similarly, there is such an overlap at phase-slip centres in
superconducting wires or at phase-slip lines in films. In this section
we show that the normal current is purely transient and contributes
only if the electric and magnetic fields change in time.

\subsection{Effective vector and scalar potentials}\label{EP}
The normal current enters the floating-kernel approximation in two 
ways: in the kinetic energy of the equation \eqref{e2} and in the 
supercurrent \eqref{e3}. In both cases ${\bf J}_{N}$ and 
${\bf A}$ appear together so that both are accounted for by a vector field
\begin{align}
{\bf A}_{\rm FK}={\bf A}+
{m^*c\over 2e^2n}{\bf J}_{N}.
\label{e10}%
\end{align}

It is advantageous to describe the vector and scalar potentials in 
a symmetric form. To this end we express the normal current
\eqref{e1} via potentials   
\begin{align}
{\bf J}_{N}=-\sigma_{N}{1\over c}{\partial{\bf A}\over\partial t}-
\sigma_{N}\nabla\phi
\label{epot1}%
\end{align}
so that 
\begin{align}
{\bf A}_{\rm FK}={\bf A}-\tau{\partial{\bf A}\over\partial t}-c\tau\nabla\phi
\label{epot2}%
\end{align}
with the characteristic time 
\begin{align}
\tau ={m^*\sigma_{N}\over 2e^2n}.
\label{e13}%                        
\end{align}

By the substitution
$\psi={\rm e}^{-ie^*\tau\phi/\hbar}~\tilde\psi$
for a homogeneous sample, $\nabla\tau=0$, we transform the 
GL equation \eqref{e2} as
\begin{align}
&{1\over 2m^*}\left(-i\hbar\nabla-{e^*\over c}{\bf A}_{\rm eff}\right)^2\tilde\psi +
\alpha\tilde\psi+
\beta\left|\tilde\psi\right|^2\tilde\psi
\nonumber\\  
&~~~=-
{\Gamma\over\sqrt{1+\gamma^2|\tilde\psi|^2}}\left({\partial\over\partial t}+
{i\over\hbar}e^*\phi_{\rm eff}+
{\gamma^2\over 2}
{\partial|\tilde\psi|^2\over\partial t}\right)\tilde\psi.
\label{efp3}%
\end{align}
with effective potentials 
\begin{eqnarray}
\phi_{\rm eff} &\equiv &\phi-\tau{\partial\phi\over\partial t},
\label{efp4}\\
{\bf A}_{\rm eff} &\equiv &{\bf A}-\tau{\partial{\bf A}\over\partial t}.
\label{efp5}%
\end{eqnarray}
The supercurrent \eqref{e5} then reads
\begin{align}
{\bf J}_{S}={e^*\over m^*}{\rm Re}~\bar{\tilde\psi}
\left(-i\hbar\nabla-{e^*\over c}{\bf A}_{\rm eff}\right)\tilde\psi
\label{efp6}
\end{align}
and other equations of the TDGL theory need not be written again
since they depend only on the amplitude $|\psi|^2=|\tilde\psi|^2$.
 
It is now clear that the system behaves as if the normal electrons are
driven by potentials $\phi$ and ${\bf A}$ while the superconducting
electrons experience effective potentials $\phi_{\rm eff}$ and 
${\bf A}_{\rm eff}$.

\subsection{Effective electric and magnetic fields}\label{EEMF}
The above effective potentials give rise to effective magnetic and electric fields. 
The transverse component of the normal current acts on the 
condensate via an effective magnetic field
\begin{align}
{\bf B}_{\rm eff}=\nabla\times{\bf A}_{\rm eff}={\bf B}-
\tau {\partial{\bf B}\over\partial t} .
\label{e12}%
\end{align}
The time variation of the normal current acts on the condensate via 
an effective electric field
\begin{align}
{\bf E'}_{\rm eff}=-{1\over c}{\partial\over\partial t}
{\bf A}_{\rm eff}-\nabla\phi_{\rm eff}={\bf E'}-
\tau {\partial{\bf E'}\over\partial t}.
\label{e15}%
\end{align}

In both effective fields the correction term vanishes in the
stationary limit. The corrections following from the floating-kernel
picture might thus become important in transient regimes or in systems
driven by oscillating fields. The AC response of the Abrikosov vortex lattice
has been discussed in \olcite{LL09}. Here we focus on vortices driven
by a steady supercurrent.

\section{Fast vortices}\label{FV}
When driven by large transport {\em DC}, vortices move fast
and the condensate is strongly reduced behind the vortex core since it 
needs a finite time to recover. Similarly, the condensate is stronger on the 
front side since the condensate needs a finite time to dissolve or move 
away. In consequence, vortices are so deformed that their interaction becomes 
anisotropic leading to transition from the triangular 
Abrikosov lattice to rows perpendicular to the current. These rows are 
known as phase-slip lines. Along the phase-slip lines vortices move with 
velocities exceeding typical velocities of normal vortices 
by one to two orders of magnitude.\cite{Berdiyorov09,Sardella09}

In this regime one can expect the floating-kernel corrections to play
an important r\^ole for two reasons. First, in the vicinity of fast
vortices time derivatives of the magnetic and electric fields are
large so that the effective magnetic field \eqref{e12} or electric
field \eqref{e15} appreciably differs from the true
field. Second, across the phase slip line there is a large normal
current since vortices are densely packed there and the condensate is
suppressed.

Above a critical current the resistivity of a superconductor changes
in dramatic steps corresponding to rearrangement of vortices inside.
Presently, this highly nonlinear response of the system can be studied
only numerically. We thus demonstrate the effect of floating-kernel
corrections using numerical simulation.

\subsection{Dimensionless equations for thin films}\label{FV1}
In films much thinner than the London penetration depth, the current
has a negligible feedback effect on the magnetic field. We thus assume
a constant homogeneous magnetic field perpendicular to the
superconducting film. According to \eqref{e12} the effective magnetic
field coincides with the true magnetic field, ${\bf B}_{\rm eff}={\bf
  B}$. It is thus convenient to take a stationary vector potential,
$\partial {\bf A} / \partial t=0$ in which case from \eqref{efp5}
follows that ${\bf A}_{\rm eff}={\bf A}$. We thus focus on the scalar
potential and floating corrections to it.

The vector potential is constant and the Maxwell equation \eqref{e9}
is unused. We must solve the TDGL equation \eqref{efp3} which in 
dimensionless units reads
\begin{align}
&\left(\nabla'-i{\bf A}'
\right)^2\psi' +\left(1-
\left|\psi'\right|^2\right)\psi'
\nonumber\\  
&~~~=
{u\over\sqrt{1+\gamma^2|\psi'|^2}}
\left({\partial\over\partial t'}+
i\phi_{\rm eff}'+{\gamma^2\over 2}
{\partial|\psi'|^2\over\partial t'}\right)\psi',
\label{efp3dl}%
\end{align}
where the order parameter $\psi'=\tilde\psi/\psi_0$ is scaled 
with the GL value $\psi_0^2=-\alpha/\beta$. The amplitude of
relaxation rate thus reads $\gamma=C\psi_0$. Dimension $x'=x/\xi$ 
scales with the GL coherence length $\xi^2=-\hbar^2/2m^*\alpha$. 
The vector potential scales with inverse distance; ${\bf A}'=
(e^*\xi/c\hbar){\bf A}$.
The time $t'=tu/\tau_{\rm GL}$ is scaled with the GL time 
$\tau_{\rm GL}=-\Gamma/\alpha=\pi\hbar/8k_{\rm B}(T_{\rm c}-T)$
and phase relaxation rate $u$  specified below from the scaling 
of the equation for the scalar potential. The scalar effective 
potential $\phi_{\rm eff}'=\phi_{\rm eff}/\varphi_0$ scales with 
the inverse time
$\varphi_0=\hbar u/e^*\tau_{\rm GL}$.  

The supercurrent \eqref{efp6} simplifies to
\begin{align}
{\bf J}_{S}'=\Real \big[\bar{\psi'}
\left(-i\nabla'-{\bf A}'\right)\psi'\big]
\label{efp6dl}
\end{align}
with ${\bf J}_{S}'=(m^*\xi/e^*\hbar\psi_0^2){\bf J}_{S}$. Equation 
\eqref{e8} for the scalar potential rescales as
\begin{align}
\nabla'^2\phi'=\nabla'\cdot{\bf J}_{S}',
\label{e8dl}%
\end{align}
where $\phi'$ is in the same units as $\phi_{\rm eff}'$.
We have used $u=\pi\hbar/4k_{\rm B}T_{\rm c}\tau$ with $\tau$ from 
Eq.~\eqref{e13} to make Eq.~\eqref{e8dl} free of numerical factors. 
The temperature dependence of $\tau_{\rm GL}$ cancels 
with the condensate density $\psi_0^2\approx n(T-T_{\rm c})/T_{\rm c}$. 
The effective potential is
\begin{align}
\phi_{\rm eff}'=\phi'-
2\left(1-{T\over T_{\rm c}}\right){\partial\over\partial t'}\phi'.
\label{potdl}%
\end{align}

In the numerical study we use values $u=6$, $\gamma=10$ and
$T=0.75~T_{\rm c}$.  The sample has size $20\xi\times 40\xi$ and
contacts of length $4\xi$ are centered at shorter sides. Our numerical
code is a modified version of that used by Berdiyorov, Milo{\v
  s}evi{\'c} and Peeters. Principally, the modification has consisted
of the addition of the normal-current term.  The reader interested in
details related to boundary conditions can find them in their paper
\olcite{Berdiyorov09} or in a closely related \olcite{Berdiyorov09b}.

\subsection{Numerical results}\label{FV2}
\begin{figure}[ht]
  \psfig{figure=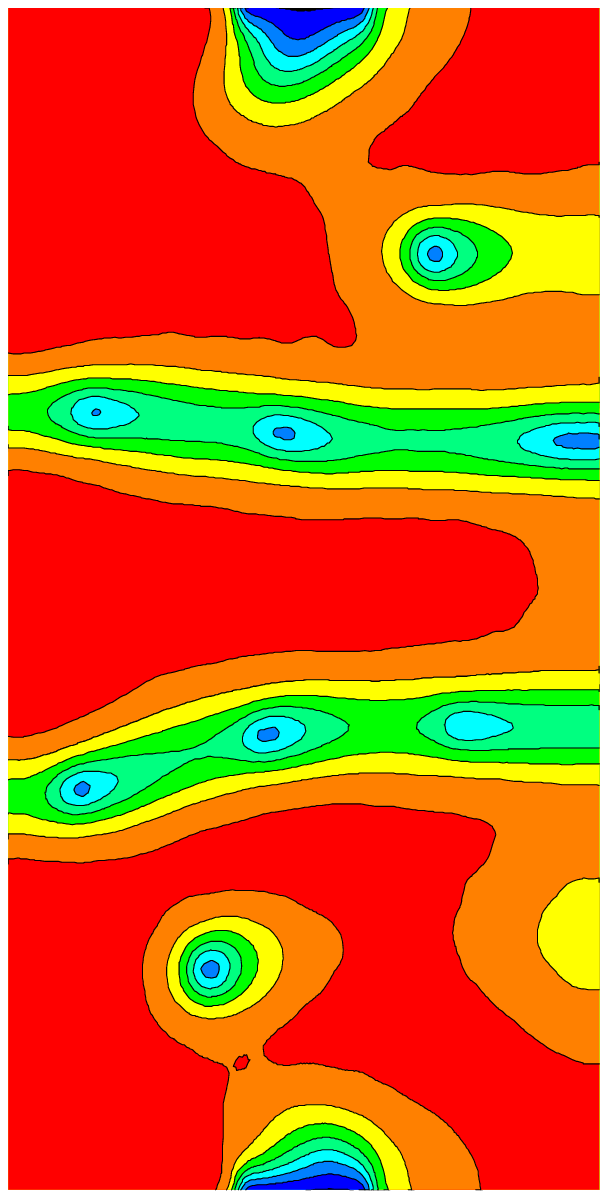,width=2.5cm,angle=0}~~~
\psfig{figure=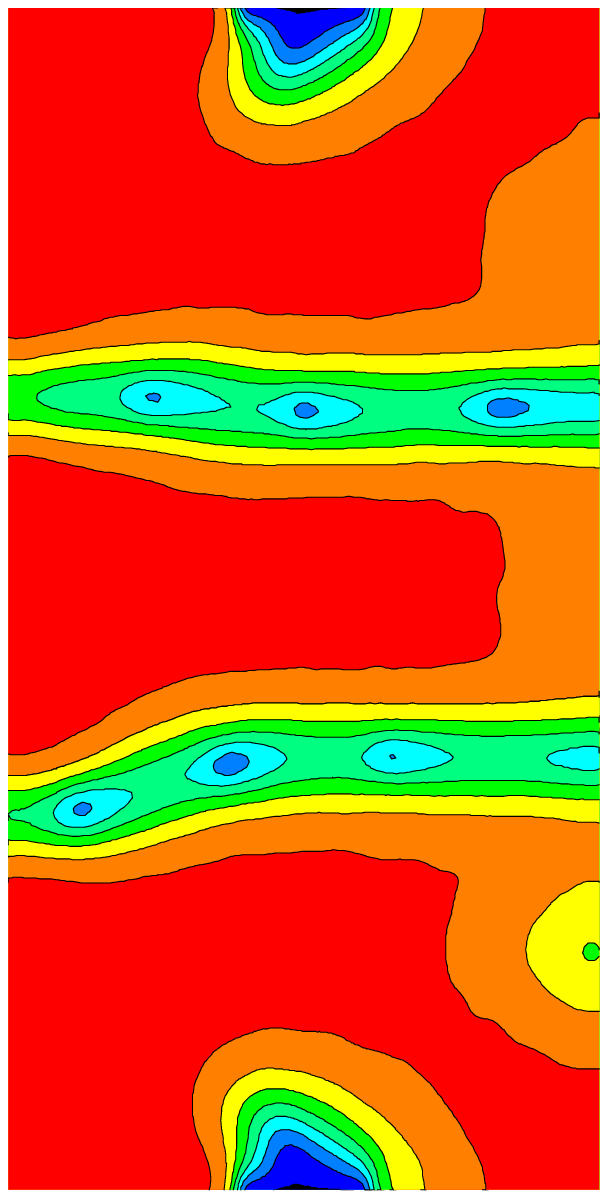,width=2.5cm,angle=0}~~~%}
\psfig{figure=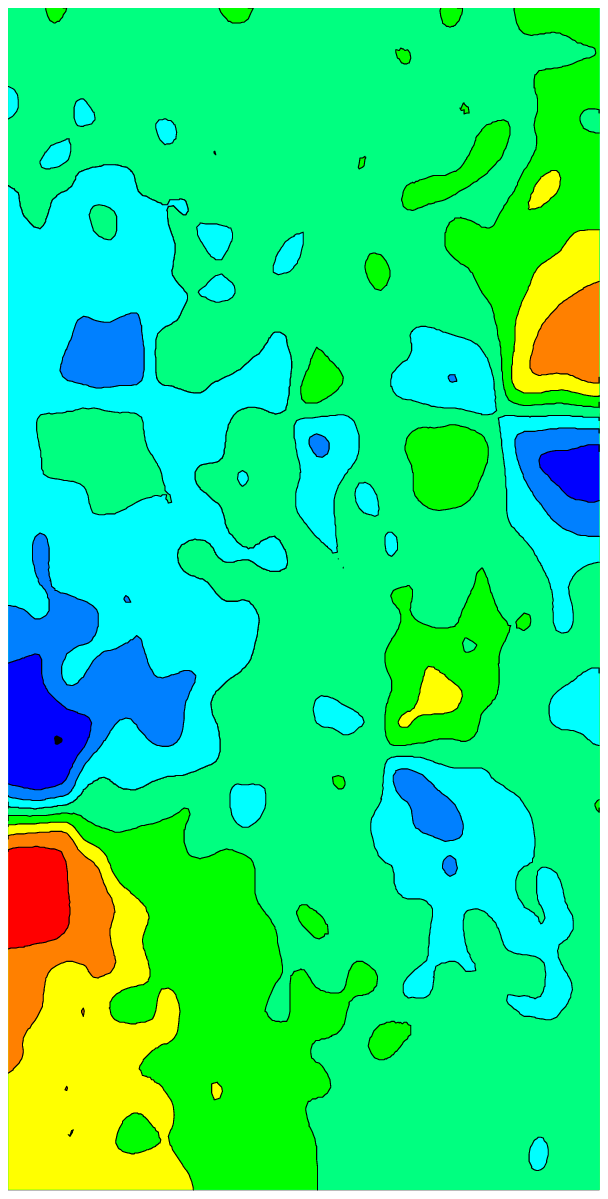,width=2.5cm,angle=0}%}
\caption{Effect of the floating-kernel correction on the condensate
  density in a 2-D sample: Low density is indicated by blue and
  maximum by red. The left-hand panel shows the TDGL theory while the
  central panel shows the floating-kernel approximation. The sample is
  in a weak magnetic field which breaks left-right symmetry. The DC is
  driven into it from normal-metal contacts creating visible
  conversion layers at horizontal edges. Near the center of the sample
  there are two phase-slip lines with kinetic vortices, closer to the
  contacts one can find isolated Abrikosov vortices. The right-hand
  panel presents the floating-kernel correction $\tau{\partial\phi /
    \partial t}$.  It is small in magnitude being a few percent of the
  scalar potential $\phi$. There is no visible contribution due to
  conversion layers, as this would correspond to the transient nature
  of the floating-kernel correction. The largest contributions (strong
  red and blue) appear near the ends of the phase-slip lines, where
  kinetic vortices achieve the highest velocity.\cite{Berdiyorov09,Sardella09} 
}
\label{fig1}
\end{figure}

\begin{figure}[ht]
  \psfig{figure=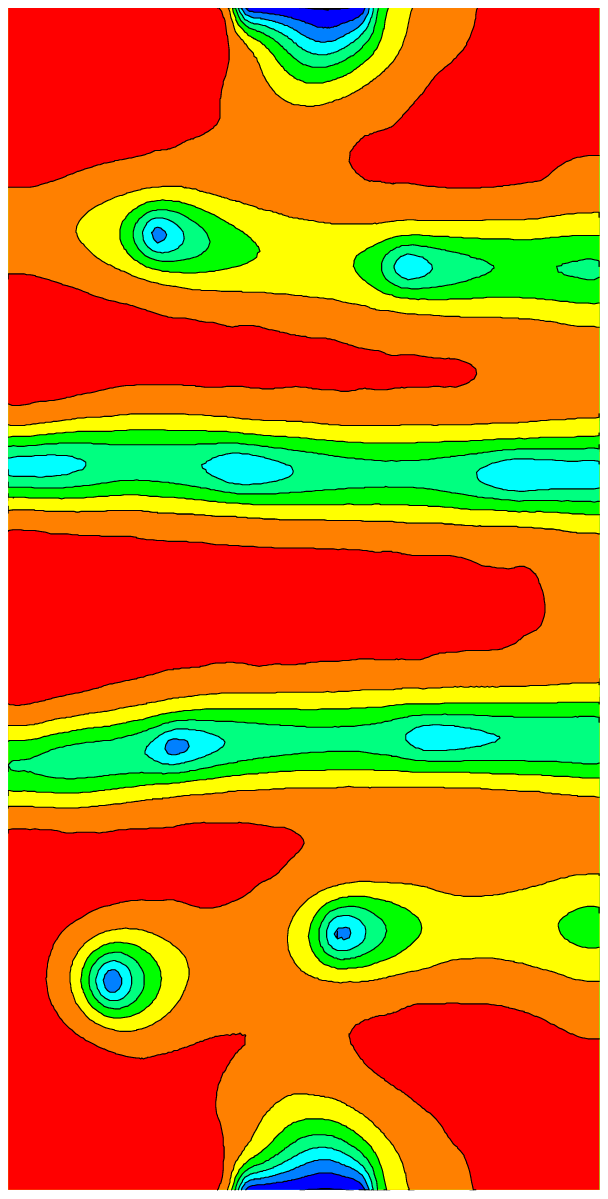,width=2.5cm,angle=0}~~~
\psfig{figure=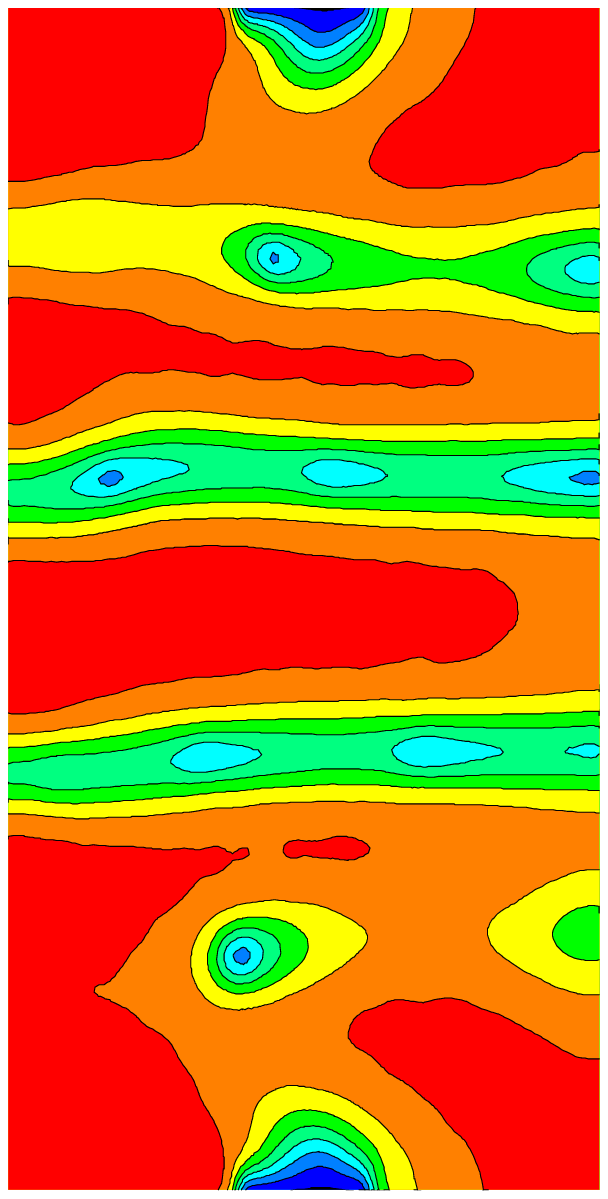,width=2.5cm,angle=0}~~~%}
\psfig{figure=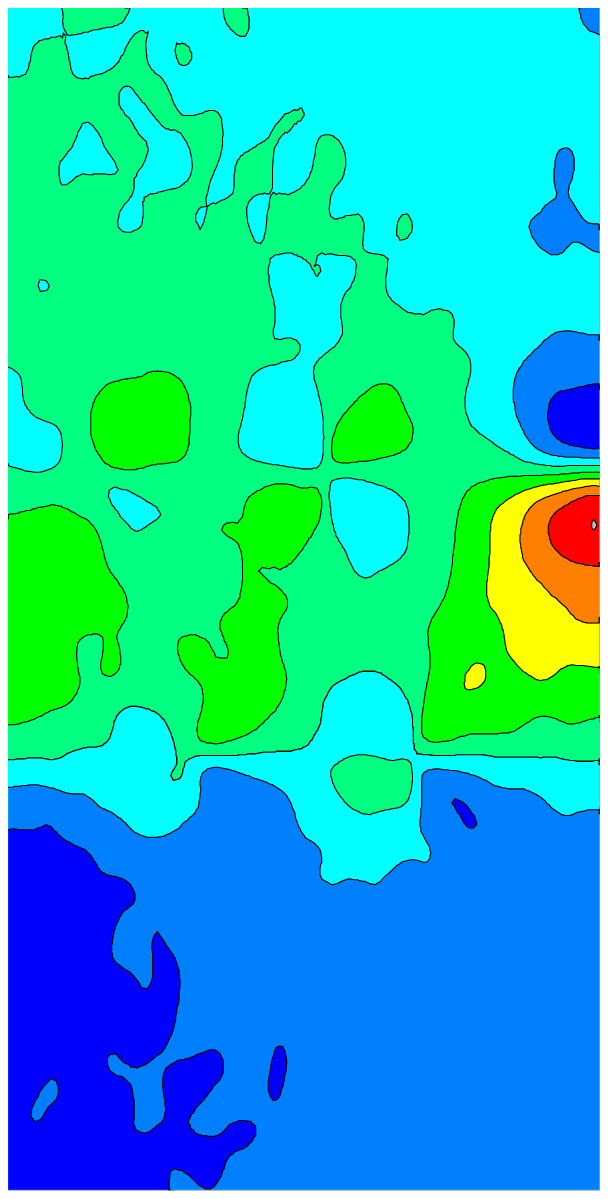,width=2.5cm,angle=0}~~~%}
\caption{As in figure \ref{fig1} but for a current increased by 15\%:
Again, there are no traces of the conversion layers in the 
floating-kernel correction. The only appreciable contribution results from
the kinetic vortex just at the sample edge. In the upper phase-slip
line one can see a four-leafed clover showing the quadrupole character
of the floating-kernel correction $\tau{\partial\phi / \partial t}$ 
of a single kinematic vortex. Similar features can also be identified 
for other kinematic vortices inside the sample.}
\label{fig2}
\end{figure}

Figures \ref{fig1} and \ref{fig2} show the order parameter in a superconducting
film in a weak magnetic field $H=0.1~H_{\rm c2}$ driven by a current from
normal contacts. Near the center of the strip the current is strong enough
to enforce a rearrangement of vortices into closely packed lines, while near 
the end points the vortices remain isolated. The strip thus sports both 
kinetic vortices in lines and isolated Abrikosov vortices. 


The left-hand panels show dynamics evaluated within the TDGL theory, the
central panels show the dynamics obtained from the floating-kernel
approximation. The overall behaviour in both approximations is very
similar. The floating-kernel correction to the scalar potential leads
to slightly different motion of vortices as one can see from their
positions after the same time of evolution has elapsed from initial
conditions of zero current and zero magnetic field.

The right-hand panel in figures \ref{fig1} and \ref{fig2} show the
floating-kernel correction to the scalar potential. It is worthy of
note that there are no visible contributions near contacts, where the
normal current is converted into the supercurrent and these two
currents have a strong overlap giving seemingly good conditions for
their interaction.  This demonstrates the transient nature of
floating-kernel corrections, as argued in \secref{EF}.

It should be noted that for the above regime the true scalar
potential has dominant contributions from conversion regions and
barely-visible features from phase-slip lines. The amplitude of 
the floating-kernel correction is only few percent on this scale. 

There are no visible contributions to the effective potential from 
isolated Abrikosov vortices. This is due to their low velocities\cite{Sardella09} 
which renders their contribution insignificant on the present scale. 

Near sample edges the floating-kernel correction to the scalar potential 
is dominated by kinematic vortices. This corresponds to the acceleration
of kinematic vortices to velocities by one order of magnitude larger near 
edges than in the interior of the sample.\cite{Sardella09,Berdiyorov09} 

In spite of a weak signal from vortices deep inside the sample, one
can see in the right-hand panel of \figref{fig2} that each vortex
brings a quadrupole floating-kernel correction to the potential. This
can be understood from the Bardeen-Stephen picture.\cite{BS65} They
have shown that a vortex moving in the horizontal direction creates an
electric field which drives the normal current through its core in the
vertical direction. The corresponding scalar potential is thus a
dipole with vertical orientation. The time derivative of this
potential due its horizontal translation has quadrupole symmetry.

\begin{figure}[ht]
  \psfig{figure=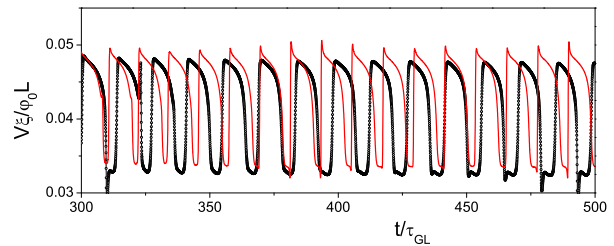,width=8cm,angle=0}
\caption{Voltage across the sample as a function of time: The 
TDGL theory (black symbols) and floating-kernel approximation (red 
line) yield similar mean voltage, 
$V_{\rm TDGL}\xi/\varphi_0L=0.0416$ and $V_{\rm FK}\xi/\varphi_0L=0.0440$.
In the floating-kernel approximation the voltage oscillates faster 
than in the TDGL case. The voltage is a difference of potentials at centres 
of contacts. % evaluated under conditions corresponding to \figref{fig1}. 
}
\label{fig3}
\end{figure}

Figure \ref{fig3} shows the voltage across a sample as a function 
of time. One can see that the mean voltages obtains by time averaging 
are rather similar; the floating-kernel gives by 6\% higher value than
the TDGL theory. As one can expect from the transient nature of the 
floating-kernel correction, a more pronounced difference appears in the 
oscillation corresponding to passage of individual kinematic vortices. The 
floating-kernel approximation as compared to the TDGL theory yields an 
increase in frequency of 17\%.  

The time dependence in \figref{fig3} was simulated under conditions
corresponding to \figref{fig1}. In \figref{fig1} one finds that 
the number of kinematic vortices in the TDGL approximation is six while
in the floating-kernel approximation there are seven. Since $7/6-1 \approx 17\%$
we conclude that vortices move with similar velocities and the different 
frequency of oscillations follows from higher density of kinematic 
vortices.

One can understand why vortex velocities are equal for both approximations
in the case of an isolated Abrikosov vortex in the following way. The vortex velocity ${\bf v}_{V}$ 
is given by the driving current density $\bar{\bf J}_{S}$ and the 
friction coefficient $\eta$ as a balance of the Lorentz and friction
force $\bar{\bf J}_{S}\times {\bf\hat z}=\Phi_0\eta{\bf v}_{V}$,
where ${\bf\hat z}$ is a direction vector either of the true magnetic
field ${\bf B}$ or of the effective field ${\bf B}_{\rm eff}$. As long as 
both fields are parallel, the vortex moves in both approximations with the
same velocity. 

Although the velocity of kinematic vortices does not obey a simple 
law of balance of forces, applicable to isolated vortices, the numerical result 
suggests that their velocities are also only negligibly influenced by 
the floating-kernel corrections. During the time evolution of the simulation,
vortex core positions were traced and their velocities calculated. It was
found that isolated vortices typically move about ten times slower than kinematic
vortices. Since velocities of kinematic vortices change strongly with
vortex positions\cite{Sardella09,Berdiyorov09} %for details,
it was not possible to compare the effect on velicities of the two 
approximations with high accuracy.

\section{Conclusions}\label{conc}
Within the floating-kernel approximation we have discussed the effect
of the normal current on the motion of the condensate. It was shown
that this effect is of transient nature and contributes only in time
dependent systems with moving vortices. The most
pronounced effects were found for kinematic vortices, which move much
faster than isolated Abrikosov vortices. Particularly strong
corrections were found near sample edges, where kinematic vortices are
accelerated, increasing their speed nearly by an order of magnitude.

\acknowledgments
Authors are grateful to Tom Lemberger who drew our attention
to the longitudinal $f$-sum rule, and to Golibjon Berdiyorov
for valuable comments on the manuscript.
This work was supported by research plans MSM 0021620834 and 
No. AVOZ10100521, by grant GA{\v C}R P204/10/0687 by DAAD and 
by Taiwan-Czech PPP project 99-2911-I-216-001.

\appendix
\section*{Appendix: Frequency sum rule}%\label{Fsumrule}
The conductivity is required to satisfy the frequency sum rule\cite{MS59,BT05}
\begin{align}
{2\over \pi}\int\limits_0^\infty 
d\omega \Real\sigma(\omega,{\bf k})={ne^2\over m},
\label{app1}
\end{align}
where $e$, $m$ and $n$ are charge, mass and density of electrons.
To satisfy this sum rule, a modification of the TDGL theory in the
spirit of the two-fluid theory is necessary. Here we show that this 
two-fluid correction implies the floating-kernel approximation 
discussed in \secref{GLFK}.

\subsection{Sum rule violation in the TDGL theory}
First we show that the standard TDGL theory leads to a conductivity which violates 
the sum rule \eqref{app1}. In the standard TDGL theory one writes the total current as 
a sum of the supercurrent and the normal current,
\begin{align}
{\bf J}_{\rm TDGL}={\bf J}_{\rm GL}+{\bf J}_{N}.
\label{app2}
\end{align}
Neglecting the Hall effect, both currents are parallel to the electric field and 
the conductivity is a scalar given by the ratio, $\sigma={\bf J}/{\bf E'}$. It thus 
has two corresponding parts
\begin{align}
\sigma_{\rm TDGL}=\sigma_{\rm GL}+\sigma_{N}.
\label{app4}
\end{align}
 
The sum rule \eqref{app1} is satisfied in the normal state
\begin{align}
{2\over \pi}\int\limits_0^\infty 
d\omega \Real \sigma_{N}(\omega,{\bf k})={ne^2\over m}.
\label{app5}
\end{align}
The superconducting component of mean Cooperon density $\bar n_{S}$, 
mass $m^*$ and charge $e^*$ has an analogous sum over frequencies
\begin{align}
{2\over \pi}\int\limits_0^\infty 
d\omega \Real \sigma_{\rm GL}(\omega,{\bf k})=
{\bar n_{S} e^{*2}\over m^*}.
\label{app6}
\end{align}
The total sum rule \eqref{app1} for $\sigma_{\rm TDGL}$ is thus violated.

\subsection{Two-fluid correction}
Assuming that formation of the condensate depletes the density of normal 
electrons, $n_{N}=n-2\bar n_{S}$, the normal 
conductivity ought to be correspondingly lowered,
\begin{align}
\sigma=\sigma_{\rm GL}+\left(1-{2\bar n_{S}\over n}\right)\sigma_{N}.
\label{app7}
\end{align}
The sum over frequencies in the left hand side of  
\eqref{app1} is then
\begin{align}
{2\over \pi}\int\limits_0^\infty 
d\omega \Real\sigma(\omega,{\bf k})=
{\bar n_{S} e^{*2}\over m^*}+
\left(1-{2\bar n_{S}\over n}\right){ne^2\over m}.
\label{app8}
\end{align}

A sum rule similar to \eqref{app8} was discussed in greater detail for
the Meissner state, where a part of the weight due to superconducting
electrons is covered by a singular term in the form of a Dirac
$\delta$ function.\cite{FG58,TF59,BT05} Avoiding the $\delta$ function
one arrives at the sum rule\cite{BT05}
\begin{align}
{2\over \pi}\int\limits_{+0}^\infty 
d\omega \Real\sigma(\omega,{\bf k})=
\left(1-{2\bar n_{S}\over n}\right){ne^2\over m}.
\label{app8mm}
\end{align}
Here we assume similar structure for the mixed state. The only 
difference is that in the presence of vortices the conductivity
$\sigma_{\rm GL}$ of superconducting electrons is finite 
even at zero frequency. Its frequency dependence is not a $\delta$ 
function but has a Drude form. 

In the pure limit, $m^*=2m$, the sum rule \eqref{app1} is satisfied.
As one can see, the sum rule \eqref{app1} is violated in the dirty
limit when $m^*\not= 2m$. This is in consequence of limitations of the
theory used to derive the floating-kernel approximation. The
derivation of \olcite{LL08} is based on the Kadanoff-Baym ansatz with
the spectral function approximated by the $\delta$ function, therefore
the renormalization of the Cooper-pair mass due to finite mean free
path is not included in this approach. Briefly, the floating-kernel
approximation is justified only in the pure limit.

\subsection{Interaction of normal current with condensate}

We note that the floating-kernel approximation discussed in 
\secref{GLFK} leads to the conductivity \eqref{app7}, with 
current \eqref{e6}. Here we approach the problem in the 
opposite way: starting from the conductivity \eqref{app7} we
arrive at the floating-kernel approximation.

Let us require that the set of TDGL equations follows
from the effective free energy\footnote{A more exact formulation
is based on the Lagrangian. We use the free energy, which
is traditional in GL theory.} $\cal F$ through
\begin{eqnarray}
\Gamma\left({\partial\over\partial t}-ie^*\phi\right)\psi&=&-
{\delta{\cal F}\over\delta\bar\psi},
\label{app9}\\
{\bf J}&=&-{\delta{\cal F}\over\delta{\bf A}}.
\label{app10}
\end{eqnarray}
Since our focus is on the spatial gradients, we neglect the non-linear 
relaxation of Kramer and Watts-Tobin. Indeed, the relaxation in the 
left hand side of \eqref{app9} results from \eqref{e2}, sending
$C\to 0$. 

In equations (\ref{app9}-\ref{app10}), the GL function $\psi$ is 
normalized to the Cooperon density as $n_{S}=|\psi|^2$ and the 
sum rule uses the value averaged over space, $\bar n_{S}=
\langle|\psi|^2\rangle$. We assume that the free energy has the
superconducting and the normal parts, ${\cal F}={\cal F}_{N}+
{\cal F}_{S}$, where the normal part is the same as in the 
normal state, therefore
\begin{equation}
{\bf J}_{N}=-{\delta{\cal F}_{N}\over\delta{\bf A}}.
\label{app10a}
\end{equation}
The GL function thus enters the superconducting part only;
\begin{equation}
{\delta{\cal F}_{N}\over\delta\bar\psi}=0.
\label{app9a}
\end{equation}

In the TDGL theory the supercurrent reads
\begin{align}
{\bf J}_{\rm GL}={e^*\over m^*}{\rm Re}~\bar\psi
\left(-i\hbar\nabla-{e^*\over c}{\bf A}\right)\psi.
\label{app3}
\end{align}
Since the vector potential ${\bf A}$ appears exclusively via the 
covariant gradient in the bracket, the current implies
that in the TDGL free energy the kinetic energy takes the familiar form
$(1/2m^*)\left|\left(-i\hbar\nabla-(e^*/c){\bf A}\right)\psi\right|^2$.

We have seen that the current \eqref{app3} with the normal current
added violates the frequency sum rule. Now we derive the kinetic
energy assuming that current includes the two-fluid correction.
According to the two-fluid conductivity~\eqref{app7}, the total
current reads
\begin{align}
{\bf J}&=\sigma_{\rm GL}{\bf E'}+\left(1-{2n_{S}\over n}\right)
\sigma_{N}{\bf E'}
\nonumber\\
&={\bf J}_{\rm GL}-{2n_{S}\over n}{\bf J}_{N}+{\bf J}_{N}
\nonumber\\
&={e^*\over m^*}{\rm Re}~\bar\psi
\left(-i\hbar\nabla-{e^*\over c}{\bf A}-
{m^*\over en}{\bf J}_{N}\right)\psi+{\bf J}_{N}.
\label{appa1}
\end{align}
We have included the correction term in the supercurrent, 
because it is proportional to the condensate density. According
to \eqref{app10} and \eqref{app9a} it thus cannot result from 
the variation of the normal free energy ${\cal F}_{N}$. 

From equations \eqref{app10}, \eqref{app10a} and \eqref{appa1} one finds
\begin{equation}
{\delta{\cal F}_{S}\over\delta{\bf A}}=-{e^*\over m^*}{\rm Re}~\bar\psi
\left(-i\hbar\nabla-{e^*\over c}{\bf A}-
{m^*\over en}{\bf J}_{N}\right)\psi.
\label{app9aa}
\end{equation}
Integrating relation \eqref{app9aa} over the vector potential one finds 
the superconducting free energy of form
\begin{align}
{\cal F}_{S}&={1\over 2m^*}\left|\left(-i\hbar\nabla-{e^*\over c}{\bf A}-
{m^*\over en}{\bf J}_{N}\right)\psi\right|^2
\nonumber\\
&~~~~~~~~~~~~~~~~~~~~~~+\alpha|\psi|^2+{1\over 2}\beta|\psi|^4.
\label{app11}%
\end{align}
Of course, the integration provides only the kinetic energy which has to be
rearranged with integration by parts into the square of covariant gradients. 
The terms independent of ${\bf A}$ represent initial conditions for the integral
and are taken from the standard GL theory. 

With free energy \eqref{app11}, equation \eqref{app9} is identical to 
the floating-kernel approximation \eqref{e2} in the $C\to 0$ limit. It should 
be noted that derivation of equation \eqref{e2} from microscopic theory was 
also carried to terms linear in ${\bf J}_{N}$, therefore additional 
terms quadratic in the normal current might appear. 

To summarize this appendix, we have shown that the TDGL theory violates 
the longitudinal $f$-sum rule. To restore the sum rule one must reduce 
the normal current which corresponds to the interaction term between the
normal current and the condensate. In this way one recovers the floating
kernel-approximation from phenomenological arguments. 

\bibliography{bose,delay2,delay3,gdr,genn,chaos,kmsr,kmsr1,kmsr2,kmsr3,kmsr4,kmsr5,kmsr6,kmsr7,micha,refer,sem1,sem2,sem3,short,spin,spin1,solid,deform,tdgl}
\end{document}